\documentclass[%
 reprint,
 amsmath,amssymb,
 aps,
]{revtex4-1}
\pdfoutput=1
\usepackage{graphicx}
\usepackage{amsmath}
\usepackage{amssymb}
\usepackage{hyperref}
\usepackage{xcolor}
\usepackage[letterpaper,left=1.5cm,right=1.5cm,top=1.6 cm,bottom=2.5cm]{geometry}

\begin{document}
\pdfoutput=1
\preprint{APS/123-QED}

\title{Bayesian estimation for quantum sensing in the absence of single-shot detection}

\author{Hossein T. Dinani$^{1,2}$}
\email[]{htdinani@gmail.com}
\author{Dominic W. Berry$^2$}
\author{Raul Gonzalez$^1$}
\author{Jeronimo R. Maze$^{1,3}$}
\author{Cristian Bonato$^{4}$}
\email[]{c.bonato@hw.ac.uk}
\affiliation{$^1$ Facultad de f\'{i}sica, Pontificia Universidad Cat\'{o}lica de Chile, Santiago 7820436, Chile}
\affiliation{$^2$Department of Physics and Astronomy, Macquarie University, Sydney, NSW 2109, Australia} 
\affiliation{$^3$Research Center for Nanotechnology and Advanced Materials CIEN-UC, Pontificia Universidad Cat\'{o}lica de Chile, Santiago 7820436, Chile}
\affiliation{$^4$Institute of Photonics and Quantum Sciences, SUPA, Heriot-Watt University, Edinburgh EH14 4AS, United Kingdom}
%\maketitle

\date{\today}

\begin{abstract}
Quantum information protocols, such as quantum error correction and quantum phase estimation, have been widely used to enhance the performance of quantum sensors. While these protocols have relied on single-shot detection, in most practical applications only  an averaged readout is available, as in the case of room-temperature sensing with the electron spin associated with a nitrogen-vacancy center in diamond. Here, we theoretically investigate the application of the quantum phase estimation algorithm for high dynamic-range magnetometry, when single-shot readout is not available. We show that, even in this case, Bayesian estimation provides a natural way to efficiently use the available information. We apply Bayesian analysis to achieve an optimized sensing protocol for estimating a time-independent magnetic field with a single electron spin associated to a nitrogen-vacancy center at room temperature and show that this protocol improves the sensitivity over previous protocols by more than a factor of $3$. Moreover, we show that an extra enhancement can be achieved by considering the timing information in the detector clicks.
\end{abstract}
\pacs{}

\maketitle
\section{Introduction}
Sensors based on individual quantum systems combine high sensitivity and spatial resolution in measuring physical quantities \cite{Degan}. Quantum information protocols, such as quantum error correction \cite{Unden}, can be used to enhance their performance and their resilience against imperfections and environmental noise. The quantum phase estimation protocol, in particular, has proven helpful for sensing over a large dynamic range \cite{Said, Cappe,Nusran,Waldherr,Hayes,Natnano,CristianPRA}. The performance of this protocol can additionally be enhanced by real-time adaptation of measurement settings based on earlier outcomes in the measurement sequence \cite{Hayes,Natnano,CristianPRA}. 

Quantum protocols have relied on single-shot measurements, which deliver detection limited only by quantum projection noise. Unfortunately, single-shot detection is not always experimentally available. In most practical situations, the classical noise is much larger than the quantum projection noise and one has to rely on the average signal from a large ensemble of experiments (``averaged" detection) \cite{Degan}. Typically, in such cases the measurement results are converted into a binary outcome by using a threshold \cite{Neumann, Nusran, Danilin}. However, it is shown that by avoiding thresholding and processing the data in a better way an enhancement can be achieved, for instance, in state detection \cite{DAnjou}. In this work, we propose a Bayesian approach to enhance the quantum phase estimation protocols in the regime of averaged-detection.

For noisy systems, various adaptive Bayesian phase estimation protocols have recently been proposed and experimentally implemented \cite{Wiebe, Paesani, Wang, Macies, Ruster}.
Here, we propose to use the standard generalized quantum phase estimation algorithm \cite{Higgins} and show that, even for the case of averaged detection, Bayes' theorem can efficiently include all information available from each measurement. An improvement in the amount of information taken into account for each measurement leads to a decrease in the number of measurements required to achieve a given estimation precision, increasing the sensitivity of the procedure. 
 This is a very general approach, applicable to different qubit systems, such as nitrogen-vacancy (NV) color centers in diamond or superconducting transmon qubits \cite{Danilin}.

In the following discussion, we focus on the sensors based on the NV center electron spin and show that the proposed protocol results in a factor of $>3$ enhancement in sensitivity compared to the protocol used in the previous work \cite{Nusran}.
\section{Magnetometry with an NV center}
 The NV center is composed of a substitutional nitrogen atom next to a vacancy in a diamond lattice. The electron spin associated with the NV center features a long coherence time and it can be initialized and read out via optical excitation and detection due to spin-dependent photo-luminescence, even at room temperature. The NV center electron spin can be used to detect a variety of physical parameters such as magnetic field \cite{Jero, Natnano}, electric field \cite{Dolde}, temperature \cite{Kucsko}, and strain \cite{Doherty}.

The NV center electron spin can measure an external time-independent (DC) magnetic field $B$ applied along its quantization axis through a Ramsey experiment, using a microwave field resonant with its ground state zero field splitting $D/(2\pi)=2.87$ GHz. 
 In this case, the spin, initialized in an equal superposition state $\left( {\left| 0 \right\rangle  + \left| 1 \right\rangle } \right)/\sqrt 2 $, evolves as $\left( {\left| 0 \right\rangle  + e^{i 2\pi f_B \tau}\left| 1 \right\rangle } \right)/\sqrt 2 $, where $\tau$ is the interaction time. Here, $f_B=\gamma B/{\left(2\pi\right)}$ is the Larmor frequency with $\gamma / (2 \pi)=28$ MHz/mT being the gyromagnetic ratio of the electron spin. By measuring in a basis rotated by an angle $\theta$ compared to the initialization basis, the probability of outcome $u=0,1$, corresponding to the spin states $\left| 0 \right\rangle $ and  $\left| 1 \right\rangle $, given $f_B$ is
\begin{equation}\label{eqpmf}
P_{m}{\left({u|f_B}\right)}=\frac{1}{2}\left({1+\left({-1}\right)^u e^{-\left({\tau/T^{\ast}_2}\right)^{2}} \cos\left({2\pi f_B \tau-\theta}\right)}\right),
\end{equation}
assuming perfect initialization/readout. The Gaussian decay factor $e^{-\left({\tau/T^{\ast}_2}\right)^{2}}$ accounts for the magnetic field fluctuations induced by a nuclear spin bath, with $T^{\ast}_2$ being the coherence time of the electron spin which is of the order of few microseconds \cite{JeroNJP}.
\subsection{Quantum phase estimation algorithm}
One goal in magnetometry is to measure the magnetic field over a large dynamic range, defined as the ratio of the maximum detectable field to the uncertainty in the field. This can be achieved, for instance, by using a protocol
 based on the quantum phase estimation algorithm~\cite{Said, Hayes, Natnano, CristianPRA}.
 The quantum phase estimation algorithm relies on a sequence of $K+1$ Ramsey measurements with exponentially decreasing interaction times $2^k \tau_0$, where $\tau_0$ is the smallest interaction time and $k=K,K-1,...,1,0$. The longest interaction time is limited by the coherence time $T^{\ast}_2$. 
 
This protocol can achieve, at best, an uncertainty in the estimate of the frequency scaling as $\propto 2^{-K}/\tau_0$. This scaling can only be reached by performing $M$ Ramseys for each interaction time, where $M$ is generally taken to scale linearly with $k$ as $M_k=G+\left(K-k\right)F$ \cite{Higgins}. Here, $G$ is the number of repetitions corresponding to the largest interaction time $2^K \tau_0$; as $k$ decreases the number of repetitions increases by $F$. The reason behind this choice for the number of repetitions is that the measurements with shorter sensing times distinguish frequencies over a wider range, so errors would have a larger impact on the variance. Therefore, errors need to be more strongly suppressed with more repetitions.
\subsection{Single-shot readout}
Previous work showed that adapting the angle $\theta$ in real time based on earlier outcomes within the measurement sequence, through a Bayesian estimation procedure, can lead to a reduction of the number of Ramsey experiments required for the sequence, providing larger sensing bandwidth~\cite{Hayes}. A sketch of this protocol, which has been experimentally demonstrated using the electron spin associated with a nitrogen-vacancy (NV) center in diamond \cite{Natnano}, is shown in Fig.~\ref{photonfig}(a). In that experiment, single-shot detection of the electron spin was achieved by resonant optical excitation of atomic-like spin-selective optical transitions~\cite{RobledoNat}. In this case, with a very high fidelity, the presence (absence) of a detector click projects the spin to the state $\left| 0 \right\rangle $ ($\left| 1 \right\rangle $) (Fig.~\ref{photonfig}(a)).

However, for the electron spin of the NV center and other spin-active defects such as the silicon-vacancy in diamond \cite{Sukachev} or other materials \cite{awschalom, nagy}, single-shot readout is available either through resonant excitation of spin-selective optical transitions at cryogenic temperatures or by nuclear spin assisted readout at room temperature \cite{Steiner, Haberle}. While room temperature operation makes nuclear-assisted readout appealing for applications, it requires a strong magnetic field and it introduces a large overhead time due to multiple readout repetitions \cite{Hopper}. 
 \subsection{Averaged readout}
\begin{figure*}[t!]
\centering
\hspace{-0.1cm}
\includegraphics[scale=0.4]{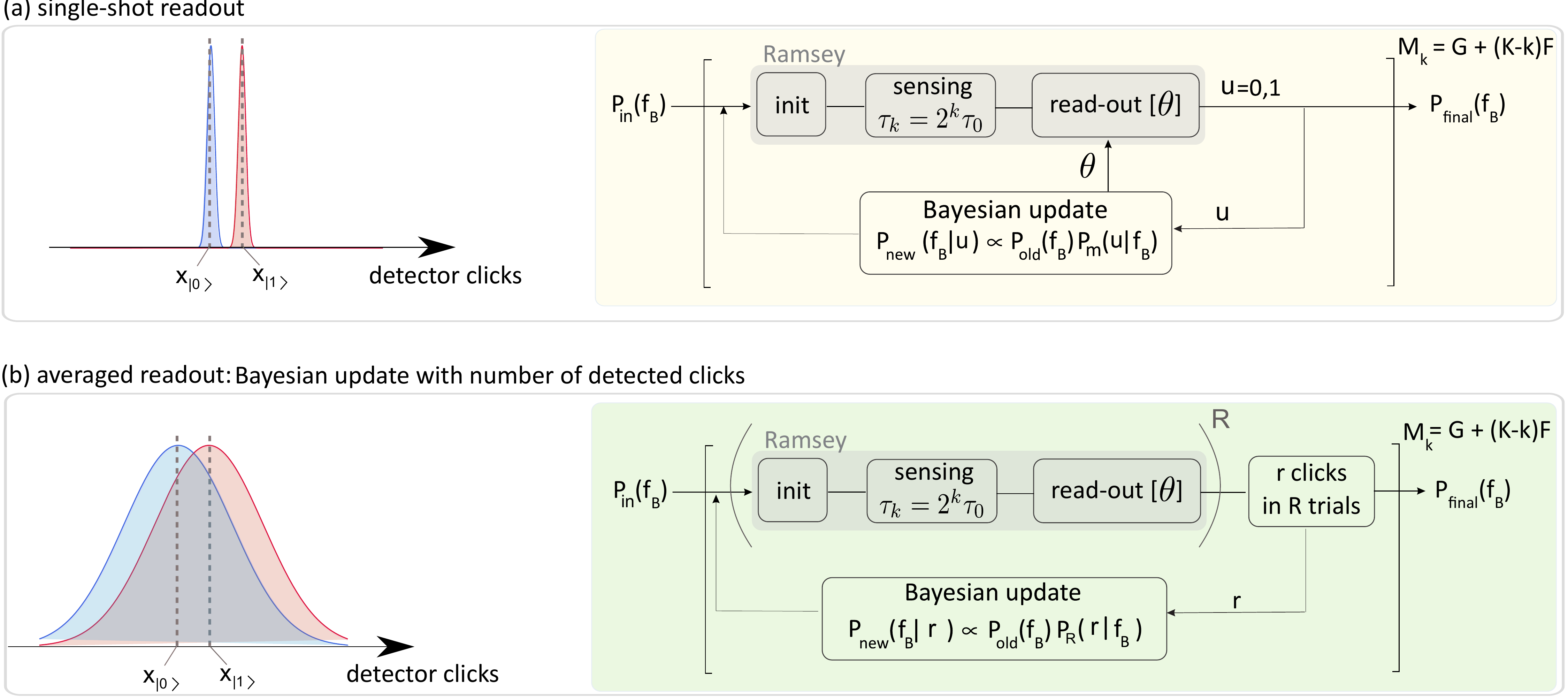}%{fig1.pdf}%{fig1_updated2.pdf}%{fig1new.eps}%{photoscheme1.png}
\caption{ A sketch of the estimation protocols. Previous work addressed implementations of the quantum phase estimation algorithm for magnetometry when single-shot readout is available, i.e., the measurements are close to projective quantum measurements (a). In this case, the outcome ($u=0,1$) of the qubit readout after each Ramsey experiment is used to update the current probability distribution $P(f_B)$ for the Larmor frequency $f_B$ through Bayes' rule. This scheme allows for real-time choice of the optimal setting for the readout basis, through the choice of $\theta$, based on the current $P(f_B)$. When single-shot readout is not available (case (b)), each individual interrogation of the qubit  does not provide sufficient information to discriminate between the spin states $\left | 0 \right \rangle$ and $\left | 1 \right \rangle$. However, a Bayesian approach, compatible with real-time adaptation of the measurement basis is still possible. } 
\label{photonfig}
\end{figure*}

At room temperature, the standard approach for detecting electron spin of the spin-active defects is based on detecting spin-dependent photo-luminescence intensity, with a contrast well below unity. In this case, due to the presence of large classical noise, the probabilities for a detector click associated to the two different spin states are quite similar and spin state discrimination cannot be achieved in a single shot, i.e., with only one Ramsey measurement (Fig.~\ref{photonfig}(b)).
 
For the electron spin of the NV center, the two spin states ($\left| 0 \right\rangle $ and $\left| 1 \right\rangle $) exhibit a difference in photo-luminescence intensity with a contrast of only roughly $0.3$, and, on average, much less than one photon is produced after each interrogation \cite{Robledo}. Typically, the experiment is repeated multiple times and the average photo-luminescence is considered. A number of approaches have been proposed to enhance the NV electron spin readout at room temperature including repetitive readout \cite{Jiang}, spin to charge conversion \cite{Shields}, analysis of photo-luminescence data \cite{Gupta}, and a statistical model for the expected distribution of measurement data \cite{Cory}. 
 \section{Threshold approach}
In previous applications of the quantum phase estimation algorithm to room-temperature sensing, each of the Ramsey measurements (which would give a binary outcome if single-shot detection were available) is replaced by an ensemble of $R$ Ramsey measurements \cite{Nusran, NusranPRB}. We call this ensemble of
$R$ measurements a ``batch". The measurement time and the angle $\theta$ are kept constant in each batch. In that approach, the spin state corresponding to the batch is determined by comparing the average photo-luminescence intensity with a threshold and retrieving a binary outcome. The threshold is normally chosen halfway between the mean probability of detecting a photon if the spin  was prepared in the states $\left| 0 \right\rangle $ and $\left| 1 \right\rangle $.

The probability distribution $P(f_B)$ for the estimated quantity $f_B$ is then updated in a sequence of $n$ batches of $R$ Ramseys based on Eq.~\eqref{eqpmf} and Bayes' theorem as
\begin{equation}
P\left(f_B|\vec{u}_n\right) \propto P\left(f_B|\vec{u}_{n-1}\right) P_{m}{\left(u_n|f_B\right)}.
\end{equation}
Here, $u_i=0,1$, $\vec{u}_n=(u_1, u_2,..., u_n)$ is the vector representing the spin states determined after each batch, and $P_{m}{(u_n|f_B)}$ is the probability of detecting the spin state $u_n$ in the $n$-th batch given the frequency $f_B$. We adopt the notation $\vec{u}_0$ being an empty vector which represents no measurement being done. The proportionality factor is just a normalization constant. We label this methodology the ``threshold approach", while in other works it has been referred to as ``majority voting''~\cite{Wang}. 

However, the accumulated photo-luminescence resulting from a batch of $R$ Ramseys includes more information than just a binary outcome. Therefore, the threshold approach may not be optimal and different methodologies taking into account all available information may provide an improvement.
\section{Bayesian protocol}
%\subsection{Updating after each measurement}
In the absence of single-shot detection it is still possible to update the probability distribution $P(f_B)$ after each measurement. We label this approach as single-measurement updating. In this case, the probability of a detector click given the frequency $f_B$ for a Ramsey experiment can be written as
\begin{eqnarray}\label{eqpdf}
%P_{D}{\left(1|f_{B}\right)} =P_{D}{\left(m_0\right)} P_{m}{\left({m_0|f_B}\right)} +P_{D}{\left(m_1\right)} P_{m}{\left({m_1|f_B}\right)},
&&P_{d}{\left(1|f_{B}\right)} =%\sum_{m=0,1}{P_{d}{\left(1|m\right)} P_{m}{\left({m|f_B}\right)}}
{P_{d}{\left(1|m_0\right)} P_{m}{\left({m_0|f_B}\right)}}\nonumber\\
&&\qquad\qquad\quad+ {P_{d}{\left(1|m_1\right)} P_{m}{\left({m_1|f_B}\right)}},
\end{eqnarray}
where $P_{d}{\left(1|m_i\right)}$, with $i=0,1$, is the probability of a detector click for the spin in the state $\left| 0 \right\rangle $ and $\left| 1 \right\rangle $, respectively.  Substituting Eq.~\eqref{eqpmf} in Eq.~\eqref{eqpdf} we obtain
\begin{equation}\label{eqpdb}
P_{d}{\left(1|f_B\right)} = \alpha \left[{1+V\cos\left({2\pi f_B \tau-\theta}\right)}\right],
\end{equation}
where $V$ is the visibility given by
\begin{equation}
V=\frac{P_{d}{\left(1|m_0\right)}-P_{d}{\left(1|m_1\right)}}{P_{d}{\left(1|m_0\right)}+P_{d}{\left(1|m_1\right)}}e^{-\left(\tau/T^{\ast}_2\right)^2},
\end{equation}
and $\alpha = \left[{P_{d}{\left(1|m_0\right)}+P_{d}{\left(1|m_1\right)}}\right]/{2}$. The probability of detecting no photon is therefore given by 
\begin{equation}\label{nphoton}
P_{d}{\left(0|f_B\right)}=1-P_{d}{\left(1|f_B\right)}.
\end{equation}
The values of $\alpha$ and $V$ are experimentally determined for the specific system in use. For the probabilities 
$P_{d}{\left(1|m_i\right)}$, with $i=0,1$, we use the mean probabilities of detecting a photon given the spin state $\left| i \right\rangle $. 

The probability $P\left(f_B\right)$ is updated based on Bayes' theorem as the following
\begin{equation}\label{recursive-d}
P (f_B|\vec{d}_n) \propto P (f_B|\vec{d}_{n-1}) P_{d}{(d_n|f_B)},
\end{equation}
where $d_n=1,0$ corresponds to detection or absence of a photon in the $n$-th measurement, and $P_{d}{\left(d_n|f_B\right)}$ is given by Eqs.~\eqref{eqpdb} and \eqref{nphoton}. The vector $\vec{d}$ represents the measurement results, i.e., $\vec{d}_n=(d_1, d_2,..., d_n)$, with $\vec{d}_0$ being an empty vector representing no measurement. Assuming no initial knowledge about the applied magnetic field, the initial probability distribution $P(f_B|\vec{d}_0)$, before any measurement is performed, is a uniform distribution in the frequency range interval $[-1/(2\tau_0),1/(2\tau_0)]$.

However, updating the probability after each measurement, in particular in an adaptive measurement, is time intensive and results in a significant overhead time. An experimentally and numerically simpler approach is to batch $R$ measurements together and use the number of $r$ detected photons from $R$ measurements. In this batching approach, for a sequence of $n$ batches with all measurements within each batch having the same settings, i.e., the same interaction time and the same angle $\theta$, the probability of $f_B$ given the measurement results $\vec{r}_n=(r_1, r_2,..., r_n)$ is updated by Eq.~\eqref{recursive-d} but replacing $d$ in the subscript with $R$ and $d_i$ in the probabilities with $r_i$.% $\vec{d}_n$ by $\vec{r}_n$, and $d_n$ by $r_n$. 
%\begin{equation}\label{recursive-r}
%P (f_B|\vec{r}_n) \propto P (f_B|\vec{r}_{n-1}) P(r_n|f_B).
%\end{equation}
%Here, $\vec{r}_n=(r_1, r_2,..., r_n)$ is the vector representing the number of photons detected in each batch in the sequence, with $\vec{r}_0$ representing no measurement being done. $P(r_n|f_B)$ is the probability of detecting $r_n$ photons in the $n$-th batch given the frequency $f_B$. Similarly, $P(f_B|\vec{r}_0)$ is a uniform distribution in the frequency range interval $[-1/(2\tau_0),1/(2\tau_0)]$.%Assuming no initial knowledge about the applied magnetic field, the initial probability distribution $P(f_B|\vec{r}_0)$, before any measurement is performed, is a uniform distribution in the frequency range interval $[-1/(2\tau_0),1/(2\tau_0)]$.

Since the probability of detecting more than one photon in one Ramsey is negligible, the probability of detecting $r$ photons in a batch of $R$ Ramseys can be written as a binomial distribution, i.e.,
\begin{equation}
P_{R}{\left({r|f_B}\right)}=\left( {\begin{array}{*{20}{c}}
{R}\\
{r}
\end{array}} \right)\left[P_{d}{\left(1|f_B\right)}\right]^r \left[1-P_{d}{\left(1|f_B\right)}\right]^{R-r},
\end{equation}
where $\left( {\begin{array}{*{20}{c}}
{R}\\
{r}
\end{array}} \right)$ is the binomial coefficient and $P_{d}{\left(1|f_B\right)}$ is given in Eq.~\eqref{eqpdb}. In the limit of large $r$ and $R$, the binomial distribution can be approximated as a Gaussian distribution for $r$ as
\begin{equation}
\label{eq:binomial_apprx}
P_{R}{\left({r|f_B}\right)} \approx \frac{1}{\sqrt{2\pi}\sigma} \exp\left \lbrace -\frac{\left[r-R P_{d}{\left(1|f_B\right)}  \right]^2}{2 \sigma^2} \right \rbrace,
\end{equation}
 with variance $\sigma^2=R P_{d}{\left(1|f_B\right)} {\left[  1 - P_{d} {\left(1|f_B\right)}\right]}$ . 
 For $R\gg1$ the variance may be approximated by replacing $P_{d}{(1|f_B)}$ with the mean value $r/R$, resulting in $\sigma^2\approx r(R-r)/R$.
 %For $R\gg1$ we approximate the probability $P_D{\left(f_B\right)}$ in the variance by $r/R$, obtaining $\sigma^2\approx r {\left(R-r\right)}/R$. % and write the distribution as a Gaussian for $P_{D}{\left(f_B\right)}$,
%\begin{equation}
%\label{eq:binomial_apprx}
%P\left({r|f_B}\right) \propto \exp\left \lbrace -\frac{\left[ \alpha \left(  1+V\cos\left({2\pi f_B \tau-\theta}\right) \right) -\mu \right]^2}{2 \sigma^2/R^2} \right \rbrace,
%P{\left({r|f_B}\right)} \propto \exp\left \lbrace -\frac{\left[P_{D}{\left(f_B\right)} -\mu \right]^2}{2 \sigma^2} \right \rbrace,
%\end{equation}
%with the mean $\mu=r/R$.
%with the mean $\mu=r/R$ and the variance $\sigma^2=P_{D}\left(f_B\right) \left[  1 - P_{D} \left(f_B\right)\right]/R$. This Gaussian is peaked around $P_{D}(f_B) \approx \mu$ and since  $r\ll R$ we can further simplify it by approximating the variance as $\sigma^2\approx \mu \left(1-\mu\right)/R = r(R-r)/R^3$. 

Equation \eqref{eq:binomial_apprx} is used in conjunction with Eq.~\eqref{recursive-d} (modified as explained above): after each batch of $R$ Ramseys, the Bayesian update consists of multiplying the current probability density $P(f_B)$ by the Gaussian in Eq.~\eqref{eq:binomial_apprx}. Figure \ref{photonfig}(b) shows a sketch of the batching protocol.  

\section{Comparison of the two approaches}
We compare the performances of the Bayesian and ``threshold" approaches by numerical simulations. 
We consider an NV center with spin coherence time $T^*_2=1.3$~$\mu$s and we select the shortest interaction time to be $\tau_0 = 12.5$~ns. We take the probabilities of a detector click, for the cases where the electron spin is initialized in the states $\left| 0 \right\rangle $ and $\left| 1 \right\rangle $, as $P_{d}{\left(1|m_0\right)}~\approx0.03$ and $P_{d}{\left(1|m_1\right)} \approx 0.02$, respectively. For these probabilities, detections are included only up to a cutoff time. We chose the cutoff time as 320~ns to maximize the signal to noise ratio (SNR) \cite{supplement}.

The estimate of the frequency, denoted by $\check f_B$, can be achieved as 
\begin{equation}
{\check f_B} = \frac{1}{2\pi \tau_0}\arg \int_{}^{} {{e^{i2\pi {f_B} \tau_0}}P\left( {{f_B}} \right)df_B},
\end{equation}
%where $\vec{r}_n=(r_1, r_2,..., r_n)$ represents the number of detected photons for each batch. 
where $P\left( {{f_B}} \right)$ is probability of $f_B$ given the measurement results obtained from Eq.~\eqref{recursive-d}. This estimate is chosen due to ease of calculations. This is an estimate normally used for periodic quantities but frequency is not periodic. Therefore, it is possible that for frequencies close to one side of the frequency cut, i.e., $\pm 1/(2\tau_0)=\pm 40$~MHz ($\tau_0=12.5$ ns), we obtain an estimate which is close to the other side of the frequency cut. To avoid this issue, in our simulations we have chosen frequencies in a slightly smaller range, i.e., $[-39,39]$ MHz. Although this results in a slight reduction in dynamic range, in the limit of accurate measurements the reduction in the dynamic range is very small. We note that this frequency range corresponds to the range of the magnetic field $\approx [-1.39,1.39]$~mT.

The accuracy in the estimates $\check f_B$ can be evaluated with the mean-square error $V_{ B}$ defined as  
\begin{equation}
%V_{ B}=\left|\left\langle e^{2i\pi\tau_0 \left(\check f_B-f_B\right)}\right\rangle\right|^{-2}-1,
{V_B} = \left\langle {\left({\check f_B} - {f_B}\right)^2} \right\rangle.%\frac{1}{M}\sum\limits_{j = 1}^M {{{\left( { \check f_{B,j}- f_{B,j}} \right)}^2}} 
\end{equation}
Here, $f_B$ is the actual frequency and the average is taken over the estimates of the frequency $\check f_B$. A fair figure of merit which takes into account all the available resources is known as the sensitivity, $\eta$, defined as the square root of the product of $V_{ B}$ and the total measurement time, $\eta={\left( {{V_B}T_{\rm tot}} \right)^{1/2}}$.%, where%\cite{Holevo}

In addition to the free evolution time, each Ramsey requires some additional overhead time for spin initialization and readout. We take the overhead time for each Ramsey to be 3 $\mu$s. Therefore, in the case of repeating each interaction time $R(G+F(K-k))$ times, the total overhead time is $T_{\rm oh}=3R(1+K)(KF+2G)/2$ in $\mu$s. The total time required to complete a full estimation sequence is given as the sum of the total evolution time and the total overhead time, $T_{\rm tot}=\left(2^{K+1}\left(G+F\right)-\left(K+2\right)F-G\right)\tau_0 R+ {T_{\rm oh}}$. %Note that $R$ repetitions are replacing a single measurement of the single-shot protocol.

\begin{figure}[t]
\centering
\vspace*{-0.7cm}\hspace*{-0.55cm}\includegraphics[scale=0.515]{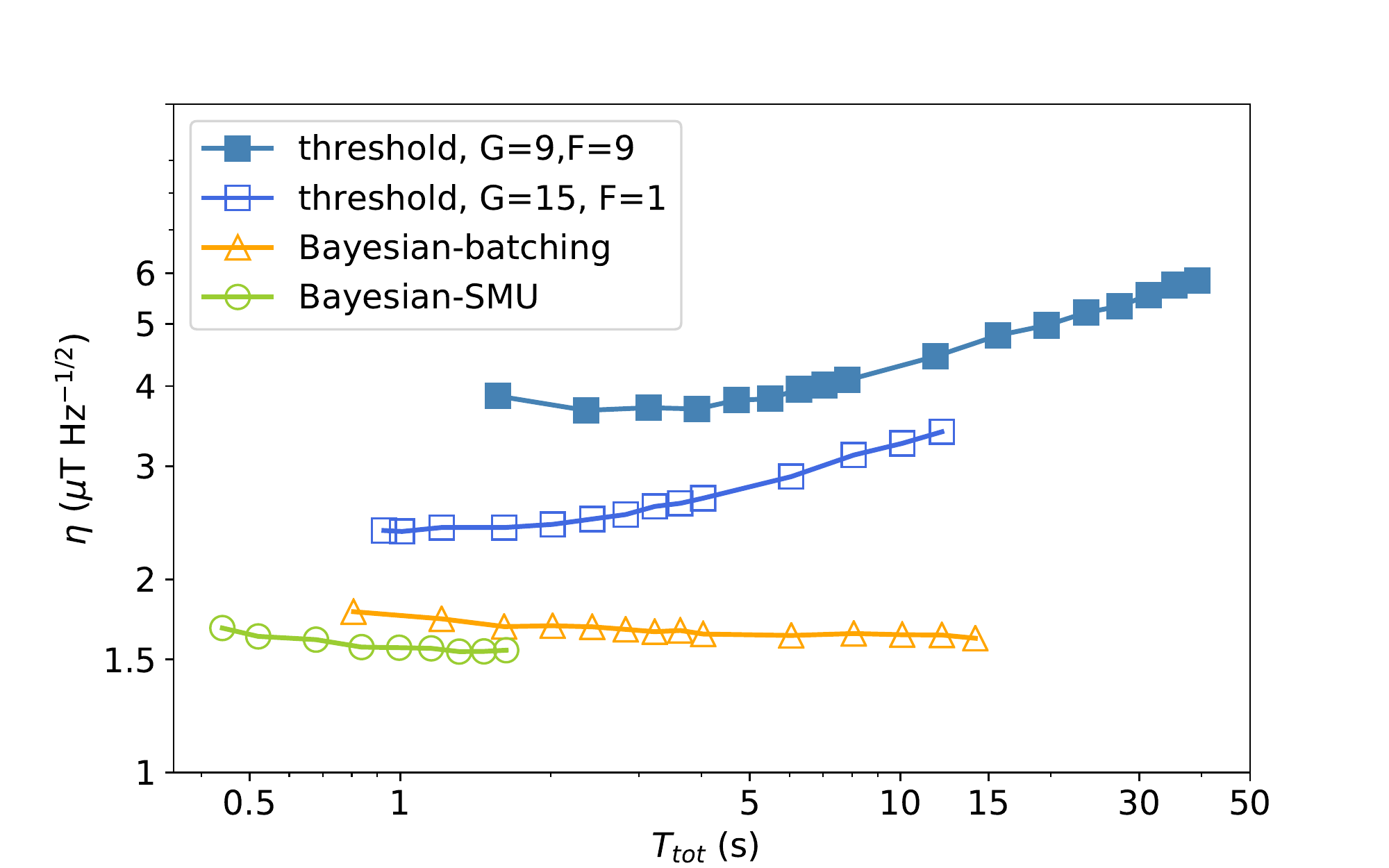}%{fig2PRB.pdf}%{p2gf.eps}%{plot1.pdf}%{plot1-nolegend.pdf}%{threshbayes.pdf}
\caption{Sensitivity $\eta$ as a function of the total measurement time, for different protocols, in the case where only ``averaged'' qubit readout is available. We consider the qubit dephasing time $T_2^* = 1.3 \, \mu$s, and we set $K=6$, i.e., the interrogation time is changed between $2^6 \tau_0$ and $\tau_0$, with $\tau_0  =12.5$~ns. Blue lines with $\blacksquare$ and $\square$ correspond to the threshold protocols for $G=F=9$ and $G=15, F=1$, respectively. Orange line with $\triangle$ shows the Bayesian batching protocol with $G=15, F=1$, while green line with $\circ$ is the Bayesian protocol with single-measurement updating (SMU) for $R=700$ and $F =1$. The data points for the threshold and batching approaches are obtained by increasing the number of repetitions $R$ of each batch of Ramsey experiments. The data points of the Bayesian with single-measurement updating is obtained by keeping $R=700$ fixed and changing $G$.}
\label{figthreshbayes}
\end{figure}

In Fig.~\ref{figthreshbayes} we have plotted the sensitivity $\eta$ as a function of the total measurement time for the threshold and Bayesian (batching and single-measurement updating) protocols. 
%We a range of values of $R$ from $10^3$ to $10^5$ repetitions (keeping $K$ fixed). 
 For the threshold protocol, assuming $G=15, F=1$, which we found to be optimal, the better sensitivity (smaller value of $\eta$) is achieved for $R \sim 2.5\times10^{3}$ ($T_{\rm tot} \approx 1$~s). For smaller values of $R$ the error, and therefore $\eta$, is significantly larger which we have not plotted to have a better scaling in the graph. The sensitivity initially improves for increasing $R$, since an increased number of repetitions enhances the discrimination of the spin state. However, for $R>2.5\times 10^3$ the reduction in measurement error given by better spin discrimination becomes less important than the increase in sensing time given by the additional repetitions.  For comparison we have also plotted the sensitivity of the threshold protocol for $G=F=9$, as in Ref.~\cite{Nusran}. For this case the optimal value of $R$ is roughly $3\times 10^3$ ($T_{\rm tot}\approx 2.35$ s). 

\begin{figure*}[t]
\centering
\includegraphics[scale=0.95]{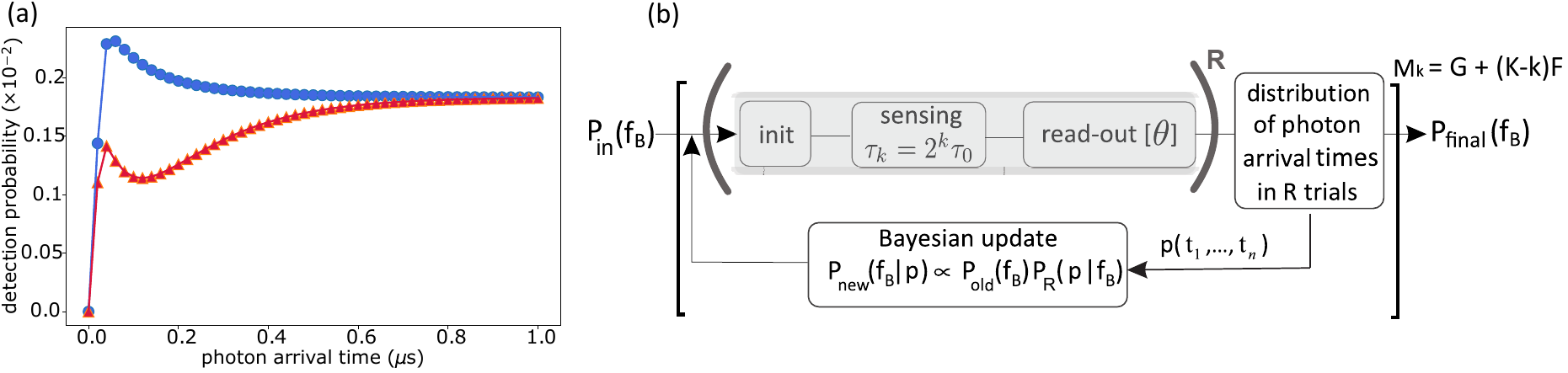}%{fig3.pdf}%{fig3new_updated.pdf}%{fig3new.eps}%{photoscheme2.png}%{photosim.pdf}%{photolumL.pdf}
\caption{(a) Optically-detected magnetic resonance experiments provide more information than the number of detected photons. When considering the arrival times of the detected photons, photon clicks corresponding to arrival times between $100$ and $200$~ns, for example, are more likely to be associated with a spin prepared in the state $\left| 0 \right\rangle $. The photo-luminescence signal is simulated based on a five-level model \cite{supplement}. The blue line with $\bullet$ (red line with $\blacktriangle$) corresponds to the spin prepared in $\left| 0 \right\rangle $ ($\left| 1 \right\rangle $) state. (b) This additional information can also be included in Bayesian estimation by updating the probability distribution for the magnetic field $P{(f_B)}$  from the distribution of photon arrival times. }
\label{photonfig2}
\end{figure*}
 The sensitivity of the Bayesian protocols appear to be always better than the one for the threshold protocol and reach a saturation for large $T_{tot}$. This is consistent with an improved way to include information from the measurement results. Whereas for the threshold protocol no additional information is used once a number of repetitions sufficient for discrimination has been reached. The Bayesian protocol uses the available information in a very efficient way. 
 
 The single-measurement updating reaches best sensitivities (smallest $\eta$) in shorter times. However, it is clear that, for large $R$, the batching approach is a good approximation for the single-measurement updating. We note,  for the single-measurement updating we found that each measurement settings, i.e., each interaction time and phase $\theta$, should be repeated $R$ times, with the optimal value found to be $R=700$. The reason is that the mean number of detected photons is small and as a result the measurements should be repeated to achieve enough information for each setting. For this case, to have a range of values of $T_{tot}$ we have kept $R=700$ fixed and have changed $G$ (with $F=1$ also fixed). Note that because single measurement updating reaches small sensitivities in shorter times, it can be used to track signals that fluctuate faster. We also note that, for all the protocols in our simulation, the readout angle $\theta$ is pre-determined, and changed in steps of $\pi/M_k$ after $R$ Ramsey measurements. 
 
The Bayesian batching (single-measurement updating) protocol achieves the sensitivity~$\eta=1.62(1.54)$ $\mu$T~Hz$^{-1/2}$, which is a factor of $\sim1.47 (1.54)$ enhancement over the best threshold protocol that we were able to find, for which $G=15, F=1, R=2.5\times10^3$~($T_{\rm tot}\approx1$~s).  The Bayesian protocol provides an even better enhancement factor of $\sim3.6 (3.7)$ as compared to the threshold protocol of Ref.~\cite{Nusran}, for which $G=9$, $F=9$, and $R=5\times10^4$ ($T_{\rm tot} = 39.2$ s).
 
The enhancement achieved depends on the photon collection efficiency and the contrast of the photo-luminescence data, i.e., the probabilities $P_{d}{\left(1|m_0\right)}$ and~$P_{d}{\left(1|m_1\right)}$. A lower contrast in the photo-luminescence could be a result of lower initial spin polarization or imperfect population transfer between the spin states. For lower contrast or photon collection efficiency the Bayesian protocol results in a lower enhancement over the threshold protocol.

\section{Including the timing information of the detector clicks}\label{arrivaltime_sec}
Optically-detected magnetic resonance experiments with NV centers provide more information than just the number $r$ of photons detected in a batch of $R$ Ramsey experiments. The arrival time of photons carries information about the spin state. The spin-dependence of photo-luminescence intensity is a result of spin-dependent inter-system crossing to metastable singlet states. In other words, since the $\left| \pm1 \right\rangle $ excited states couple more strongly to the long-lived singlet states than the $\left| 0 \right\rangle $ state, the $\left| \pm1 \right\rangle $ state exhibits reduced photo-luminescence compared to the $\left| 0 \right\rangle $ state during the first few hundred nanoseconds after optical excitation \cite{supplement}. 

This is evidenced in Fig.~\ref{photonfig2}(a), which shows the photo-luminescence signal, when either $\left| 0 \right\rangle $ or $\left| 1 \right\rangle $ states are prepared, as a function of  time after optical excitation. For the case plotted, the spin difference in the photo-luminescence signal is significant up to $\sim 700$~ns. 
For example, a photon arriving at $60$~ns is more likely to correspond to the spin being in the $\left| 0 \right\rangle $ state. No such information is available on longer timescales: a photon is equally likely to be detected at $1$~$\mu$s for the spin in the states $\left| 0 \right\rangle $ or $\left| 1 \right\rangle $.

We will now discuss how this additional information contained in the arrival time of the photons can be included by Bayesian estimation, and quantify the advantage in terms of sensitivity. The probability to detect a photon at time $t_i$ given $f_B$, $P^{(i)}_{d}{\left(1|f_B\right)}$, can be written in terms of the probability of detecting a photon at time $t_i$ given the spin state $\left| m \right\rangle $, $P^{(i)}_{d}\left(1|{m}\right)$, and the probability of the spin state $\left| m \right\rangle $ given $f_B$, $P_{m}{\left({m|f_B}\right)}$, as the following
\begin{eqnarray}
P^{(i)}_{d}{\left(1|f_B\right)}&=&P^{(i)}_{d}{\left(1|{m_0}\right)}P_{m}{\left({m_0|f_B}\right)}\\ \nonumber
&&+P^{(i)}_{d}{\left(1|m_1\right)}P_{m}{\left({m_1|f_B}\right)}.
\end{eqnarray}

In the Bayesian protocol, we considered the mean number of photons detected up to only $t_{\rm opt}=320$ ns. This is the time interval which maximizes the SNR. However, the differential photo-luminescence goes beyond this optimal time. For the example shown in Fig.~\ref{photonfig2}(a) the difference is significant up to $\sim 700$~ns. Considering the arrival time of photons, it is advantageous to take into account the photo-luminescence data beyond $320$ ns and up to time $700$ ns. We discretize the time interval $\left[{0, 700}\right]$ ns to time bins. To simplify the numerical calculations we have only considered 4 time bins \cite{supplement}. For the time bin $\Delta t_i$, $P^{(i)}_{d}{\left(1|m\right)}$ is the mean probability of detecting a photon for this time bin given the spin state $\left| m \right\rangle $, obtained from the photo-luminescence simulations shown in Fig.~\ref{photonfig2}(a). 

To take into account the timing information, in the single-measurement updating case, if a photon is detected in the time bin $\Delta t_i$ we update the probability distribution $P(f_B)$ with Eq.~\eqref{eqpdb}, replacing $P_{d}{(1|m_i)}$ with $P^{(i)}_{d}{\left(1|m_i\right)}$. If no photon is detected, we update the probability $P(f_B)$ with Eq.~\eqref{nphoton}, taking into account that the probabilities $P_{d}{(1|m_0)}$ and $P_{d}{(1|m_1)}$ are the sum of the probabilities of all the time bins.
 
In the batching approach, for a batch of $R$ Ramseys with the same measurement settings, the probability of detecting $r_1$ photons in the time bin $\Delta t_1$, $r_2$ photons in the time bin $\Delta t_2$,..., and $r_4$ photons in the time bin $\Delta t_4$ is a multinomial distribution which can be approximated as a multivariate Gaussian distribution, i.e.,
\begin{equation}\label{eq-multivarg}
%P({{{\vec {r}}}}_{}|f_B) \approx \prod\limits_{i = 1}^{4} {\exp \left[ { - \frac{{{{\left( {{P^{(i)}_D{\left(f_B\right)}} - {\mu _i}} \right)}^2}}}{{2{\mu _i}\left( {1 - {\mu _i}} \right)/R}}} \right]}.
P_{R}{({{{\vec {r}}}}_{}|f_B)} \approx \prod\limits_{i = 1}^{4} {\frac{1}{\sqrt{2\pi}\sigma_i}\exp \left[ { - \frac{{{{\left( {{r_i-RP^{(i)}_{d}{\left(1|f_B\right)}} } \right)}^2}}}{{2\sigma^2_i}}} \right]}.
\end{equation}
Here, $\vec{r}$ is the vector of detected photons in the time bins, $\vec{r}=(r_1, r_2, r_3, r_4)$, $P^{(i)}_{d}{(1|f_B)}$ is given by Eq.~\eqref{eqpdb}, replacing $P_{d}{(1|m_i)}$ with $P^{(i)}_{d}{(1|m_i)}$. Using the same approximation as in Eq.~\eqref{eq:binomial_apprx} for the variance we have $\sigma^2_i\approx r_i(R-r_i)/R$.

\begin{figure}[t]
\centering
\vspace*{-0.5cm}\hspace*{-0.6cm}\includegraphics[scale=0.515]{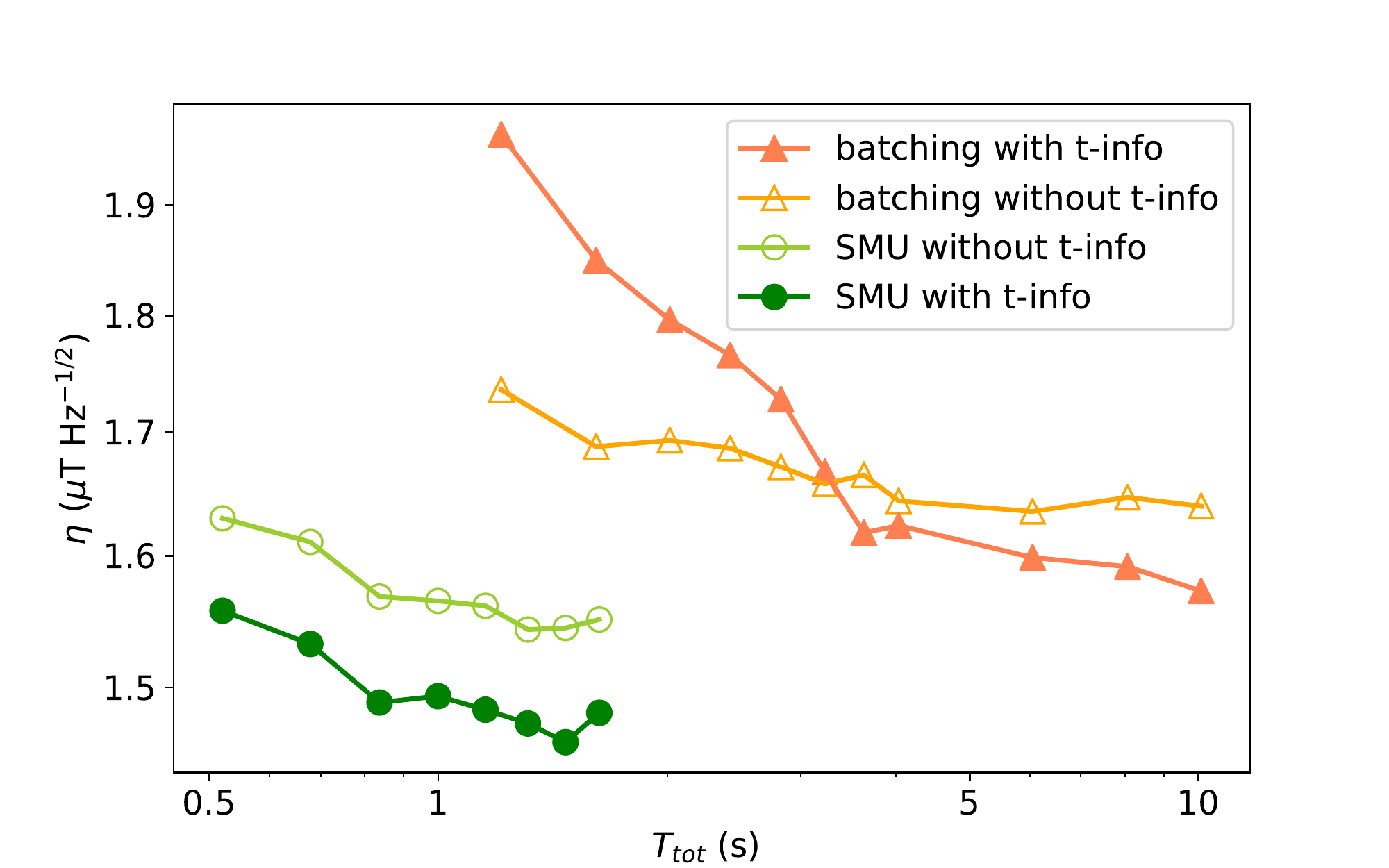}%{fig4PRB.pdf}%{ptba.eps}%{plot2b.pdf}%{plot2-nolegend.pdf}
\caption{Sensitivity $\eta$ versus total measurement time. The orange lines with $\blacktriangle$ and $\triangle$ show the batching approach with and without taking the arrival time information (t-info) of photons into account, respectively. For these lines we have set $K=6$ and $G=15, F=1$ while varying $R$, the number of repetitions. The green lines with $\bullet$ and $\circ$ show the single-measurement updating (SMU) approach with and without the arrival time information, respectively. In this case $R=700$ and $F=1$ are kept fixed and $G$ is varied.}
\label{figenh}
\end{figure}
%Taking the arrival time of photons into account and using a Bayesian protocol we obtain an extra enhancement.

In Fig.~\ref{figenh} we have compared the Bayesian protocols with and without considering the arrival time information of photons. This figure shows that including the timing information in the batching approach only results in a slight enhancement for large values of $R$, corresponding to large $T_{tot}$. The reason is for small values of $R$ ($R<10^4$) the number of detected photons in some of the time bins is small and therefore the Gaussian is not a good approximation. On the other hand, including the timing informaiton in the single-measurement updating case  results in an enhancement even for small $T_{tot}$. In this case the enhancement achieved is up to $\sim10\%$ over the corresponding Bayesian protocol without the timing information. 
%Based on our simulations, we expect to achieve about the same enhancement even considering a larger number of time bins. 

While all the simulations above consider a non-adaptive protocol, where the controlled phase is deterministically updated at each step according to a pre-determined rule, recent work has shown that real-time adjustment of the controlled phase based on previous measurement outcomes can provide advantages in terms of measurement bandwidth. We note that our Bayesian approach is compatible with real-time adaptation of the controlled phase. However, for simplicity we have only presented the nonadaptive results. 
\section{Conclusions}
We analyzed how the quantum phase estimation algorithm can be used efficiently for high dynamic range sensing with a single qubit, in the case where single-shot readout is not available, for example in room-temperature magnetometry with the electron spin of an NV center in diamond. Our results show that using Bayesian estimation to update the probability after every single measurement results in enhancement of the sensitivity over the threshold protocol. We also showed that  batching the measurements together and using the number of detected photons in the Bayesian updating is a good approximation of updating the probability after every single measurement. The batching approach is easier to implement experimentally.

An important figure of merit is the ratio between the range of the magnetic field and the sensitivity. Our proposed Bayesian protocol achieves $B_{\rm max}/\eta\approx 859(902)$ Hz$^{1/2}$, for batching (single-measurement updating), which is a factor of $\sim 3.6 (3.7)$ enhancement over the threshold protocol used in previous work. Moreover, we showed that using additional information on the arrival time of the detected photons can further enhance the sensitivity up to $\sim 10\%$.
 
The Bayesian protocol proposed here could also be useful in sensing applications with other qubit systems, such as superconducting transmon qubit \cite{Danilin}, in cases where only an averaged readout is available. Moreover, our findings could be extended beyond DC magnetometry to the characterization of the qubit environment  through dynamical decoupling \cite{Zwick, Schmitt, Plenio}. Moreover, as the single shot readout of the nuclear spin state at room temperature is also achieved through repetitive readout of the electron spin and comparing with a threshold \cite{Neumann, Wrachtrup16} this protocol may also be useful in enhancing the readout of the nuclear spin.

\section{Acknowledgment}
The authors thank Machiel Blok, Raffaele Santagati, Antonio Gentile, and Benjamin D'Anjou for helpful discussions. H.T.D.~acknowledges support from the Fondecyt-postdoctorado grant No.~3170922. D.W.B.~was funded by Discovery Project No.~DP160102426 and DP190102633. J.R.M.~acknowledges support from Conicyt-Fondecyt Grant No.~1180673, AFOSR Grant No.~FA9550-18-1-0438 and FA9550-18-1-0513, and Fondef Grant No.~ID16I10214. C.B.~acknowledges support by the Engineering and Physical Sciences Research Council through grant EP/P019803/1, the Royal Society (RG170203) and the Carnegie Research Trust (RIG007503).

%\end{document}

%%%%%%%%%% Merge with supplemental materials %%%%%%%%%%
\pagebreak
\widetext
\begin{center}
\textbf{\large Supplemental Materials: Bayesian estimation for quantum sensing in the absence of single-shot detection}\label{SM}
\end{center}
%%%%%%%%%% Merge with supplemental materials %%%%%%%%%%
%%%%%%%%%% Prefix a "S" to all equations, figures, tables and reset the counter %%%%%%%%%%
\setcounter{equation}{0}
\setcounter{figure}{0}
\setcounter{table}{0}
\setcounter{section}{0}
\makeatletter
\renewcommand{\theequation}{S\arabic{equation}}
\renewcommand{\thefigure}{S\arabic{figure}}
\renewcommand{\bibnumfmt}[1]{[S#1]}
\renewcommand{\citenumfont}[1]{S#1}
%%%%%%%%%% Prefix a "S" to all equations, figures, tables and reset the counter %%%%%%%%%%

\section*{NV center photo-luminescence}
\begin{figure*}[t!]
\label{fig_SM}
\centering
\includegraphics[scale=1.1]{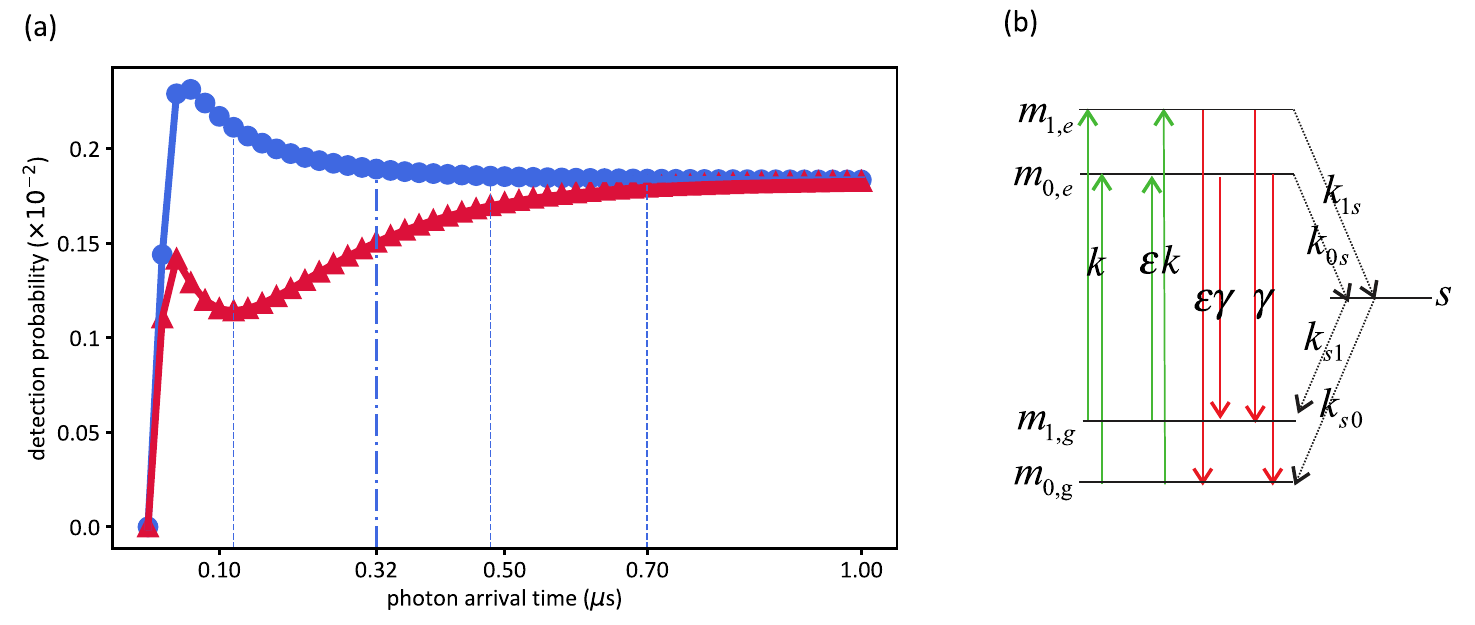}%{fig_sup.pdf}%{photolum_updated.pdf}%{photolum_sup.pdf}\,%{photolum_math2.eps}\,\qquad\quad
\caption{(a) Photo-luminescence (number of detected photons per 20 ns) as a function of the time after the excitation laser. The blue line, with $\bullet$, shows detected photons for the spin initialized optically to $m=0$ (with 0.85 polarization). The red line, with $\blacktriangle$, shows detected photons after implementing a $\pi$ pulse to polarize the spin to $m=1$ state. The vertical lines show the 4 time bins used in the protocol that includes the arrival time of photons, described in Section \ref{arrivaltime_sec}. The dot-dashed vertical line at $t_{\rm opt}=0.32$ $\mu$s  shows the optimal time which maximizes the SNR. The mean number of photons up to $t_{\rm opt}$ are used in the Bayesian and the threshold protocol. (b) The energy levels and allowed transitions of the NV center. Solid, green and red lines, show the allowed optical transitions. Dashed lines show non-radiative transitions to and from the singlet state. }
\label{photodiag}
\end{figure*}

The photo-luminescence data, shown in Fig.~\ref{photodiag}(a), and in Fig.~3(a) of the main text, are obtained by solving the rate equations for the populations of the NV spin states, considering a five level model \cite{RobledoS}. Introducing the population vector $P^{T} = \left( {{{p_{0g}}}\,\, {{p_{1g}}}\,\,{{p_s}}\,\,{{p_{0e}}}\,\,{{p_{1e}}} } \right)$, the population dynamics of the states can be written as $\dot P = M\,P$ where the matrix $M$ is given as
\begin{equation}
M = \left( {\begin{array}{*{20}{c}}
  { - k - \varepsilon k}&0&{{k_{s0}}}&\gamma &{\varepsilon \gamma } \\ 
  0&{ - k - \varepsilon k}&{{k_{s1}}}&{\varepsilon \gamma }&\gamma  \\ 
  0&0&{ - {k_{s1}} - {k_{s0}}}&{{k_{0s}}}&{{k_{1s}}} \\ 
  k&{\varepsilon k}&0&{ - {k_{0s}} - \gamma  - \varepsilon \gamma }&0 \\ 
  {\varepsilon k}&k&0&0&{ - {k_{1s}} - \gamma  - \varepsilon \gamma } 
\end{array}} \right).
\end{equation}

Figure \ref{photodiag}(b) shows a diagram of the five energy levels of the NV center and the possible transitions. In our simulations we have adopted the decay rates from the NV3 sample in Ref.~\cite{GuptaS}:\\
$\bullet$ The decay rate from the excited state to the ground state is $\gamma=66.08$ MHz. \\
$\bullet$ The decay rate from the excited states $m=0$ and $m=1$ to the singlet state labeled by $\it s$ are $k_{0s}=11.1$ MHz and $k_{1s}=91.9$ MHz, respectively.\\
$\bullet$ The decay rates from the singlet to the $m=0$ and $m=1$ ground states are $k_{s0}=4.9$ MHz and     $k_{s1}=2.03$ MHz, respectively. \\
$\bullet$ The non-spin-conserving transitions are taken to be zero, i.e., $\varepsilon=0$. \\
$\bullet$ The excitation rate $k$ is taken to be $20$ MHz.

Note that, for the chosen parameters, the difference photo-luminescence between the $m=0$ and $m=1$ states is significant up to about $700$ ns (see Fig.~\ref{photodiag}(a)). However, the optimal detection time interval for the Bayesian and the threshold protocol, described in the main text, is the time interval which maximizes the signal to noise ratio (SNR), defined as 
\begin{equation}
{\rm SNR}=\frac{{N_0-N_1}}{\sqrt{({N_0+N_1})/{2}}}.
\end{equation}
 Here, $N_m$ is the number of photons detected if the spin is prepared in $m$ state. For the photo-luminescence data shown in Fig.~\ref{photodiag}(a) we have $t_{\rm opt}=320$ ns (shown by the dot-dashed vertical line).

The number of photons detected in the detection interval $t_{\rm det}$ is proportional to $\int_0^{{t_{{\mathop{\rm det}} }}} {\gamma \left[ {{p_{0e}}(t) + {p_{1e}}(t)} \right]} \,dt$ where the proportionality factor is determined by the collection efficiency, which is taken to be 1$\%$. We take the initial spin polarization as the steady state solution to the rate equations after letting the populations in the excited and singlet state relax to the ground state in the absence of optical excitation. For the decay and excitation rates we have used, we achieve $0.85$ and $0.15$ initial polarization for the $m=0$ and $m=1$ states,  respectively. We have considered the initial population before the optical excitation to be equally distributed in the ground state. The photo-luminescence for the $m=1$ state is obtained after applying a $\pi$ microwave pulse which completely flips the polarization between the $m=0$ and $m=1$ states in the ground state. 

Note that the contrast and SNR of photo-luminescence data, and therefore the sensitivity of the magnetometry protocol, vary with the optical excitation rate which in turn depends on the power of the laser.

\end{document}